\documentclass[%
reprint,
showpacs,preprintnumbers,
 amsmath,amssymb,
 aps,
 prl,
 longbibliography,
 lengthcheck,%
]{revtex4-1}

\usepackage{graphicx}
\usepackage{epstopdf}
\usepackage{dcolumn}
\usepackage{bm}
\usepackage{hyperref}

\begin{document}

\title{Neutron, electron and X-ray scattering investigation of Cr$_{1-x}$V$_{x}$ near Quantum Criticality.}

\author{D. A. Sokolov$^{1}$, M. C. Aronson$^{1,2}$, L. Wu$^{1}$, Y. Zhu$^{1,2}$, C. Nelson$^{1}$, J. F. Mansfield$^{3}$, K. Sun$^{3}$, R. Erwin$^{4}$, J. W. Lynn$^{4}$, M. Lumsden$^{5}$, and S. E. Nagler$^{5}$.}

\address{$^1$ Brookhaven National Laboratory, Upton, NY 11973, USA\\
$^2$ Department of Physics and Astronomy, Stony Brook University, New York 11794, USA\\
$^3$ Department of Materials Science and Engineering, The University of Michigan, Ann Arbor, MI 48090, USA\\
$^4$ NIST Center for Neutron Research, NIST, Gaithersburg, MD 20899, USA
$^5$ Quantum Condensed Matter Division, Oak Ridge National Laboratory, Oak Ridge, TN 37831, USA
}

\date{\today}

\begin{abstract}
The weakness of electron-electron correlations in the itinerant antiferromagnet Cr doped with V has long been considered the reason that neither new collective electronic states or even non Fermi liquid behaviour are observed when antiferromagnetism in Cr$_{1-x}$V$_{x}$ is suppressed to zero temperature. We present the results of neutron and electron diffraction measurements of several lightly doped single crystals of Cr$_{1-x}$V$_{x}$ in which the archtypal spin density wave instability is progressively suppressed as the V content increases, freeing the nesting-prone Fermi surface for a new striped charge instability that occurs at x$_{c}$=0.037. This novel nesting driven instability relieves the entropy accumulation associated with the suppression of the spin density wave and avoids the formation of a quantum critical point by stabilising a new type of charge order at temperatures in excess of 400 K. Restructuring of the Fermi surface near quantum critical points is a feature found in materials as diverse as heavy fermions, high temperature copper oxide superconductors and now even elemental metals such as Cr.
\end{abstract}

\vspace{-0.8cm}

\maketitle
\subsection{Introduction}
A quantum critical point (QCP) is generated when matter undergoes a continuous transition from one state to another at zero temperature under the action of a nonthermal control parameter $\Gamma$$_{c}$ such as external pressure, magnetic field, or chemical doping. The suppression of electronic order to zero temperature would ultimately lead to an entropy singularity at a QCP\cite{rost2009}, but the violation of the third law of thermodynamics can be  prevented by the formation of novel collective phases, including unconventional superconductivity in heavy fermion materials such as CeCu$_{2}$Si$_{2}$\cite{assmus1984}, CePd$_{2}$Si$_{2}$\cite{mathur1998}, and UGe$_{2}$\cite{saxena2000}, and  `electron nematicity' in Sr$_{3}$Ru$_{2}$O$_{7}$\cite{borzi2007}. There is much interest in identifying new compounds that are naturally near QCPs. Materials of this class are important for their exquisite sensitivity to external fields, and the most extremal response occurs when the ordered state is tuned just to the verge of instability, i.e. at a QCP. A primary obstacle to designing materials based on this functionality is the inherent difficulty in connecting the physics of a parent material with that of the modified QC material where the underlying electronic structure is often vastly different.

There are few systems where a Fermi surface (FS) has been tracked from the ordered state to the QCP, and where the underlying electronic structure is simple enough to unambiguously identify the signatures of QC. A simple electronic structure, a well established experimental and theoretical understanding of the FS, and the itinerant nature of the $\emph{d}$ electrons all make Cr an ideal material for studies of QC. Elemental Cr undergoes a spin density wave (SDW) transition to antiferromagnetic (AF) order at 311 K that is driven by the nesting of large, parallel sections of the FS~\cite{fawcett1988}. High pressures suppress the related charge density wave (CDW), which disappears continuously at the critical pressure P$_{C}$$\sim$10 GPa\cite{jaramillo2009}. Strong charge fluctuations transverse to the wave vector of the modulation were reported to destroy the Bardeen-Cooper-Schrieffer state of the SDW, whereas the FS nesting was even improved under pressure. Chemical doping introduces electrons or holes that can change the FS volume and reduce the SDW nesting~\cite{fawcett1988}. We present here a comprehensive experimental study of  Cr$_{1-x}$V$_{x}$ that uses V doping to tune the SDW to a QCP. Neutron diffraction experiments that probe the static magnetic moment and electron diffraction measurements that probe the charge order connect variations in the order parameter and the wave vector of the coupled SDW/CDW instabilities to the electronic structure both near and far from the QCP.

\subsection{Experimental}
Single crystals of Cr$_{1-x}$V$_{x}$, x=0.0, 0.02, and 0.037, were grown
by the arc zone melting method at the Materials Preparation Center
at the Ames National Laboratory. Neutron diffraction measurements were
carried out on these single crystals of Cr$_{1-x}$V$_{x}$ using the
BT-9 triple-axis spectrometer at the National Institute for
Standards and Technology Center for Neutron Research, using a fixed
final energy E$_{F}$=14.7 meV. The measurements were performed using a
40'-47'-42'-82' configuration. Data were collected near
the (1,0,0) and (0,1,0) reciprocal lattice positions in the (HK0) plane.
Some of our early results were obtained using the
HB3 triple-axis spectrometer at the High Flux Isotope Reactor at Oak
Ridge National Laboratory. X-ray diffraction measurements were performed using
X-16B instrument at the National Synchrotron Light Source at the Brookhaven National Laboratory.
Electron microprobe measurements were made at a 20 kV accelerating voltage on a CAMECA
SX100 electron microprobe analyzer using Cr and V X-ray lines
calibrated with Cr and V standards. TEM measurements were carried
out on single crystals of Cr$_{1-x}$V$_{x}$, x=0, 0.02, and 0.37 using JEOL 2010F
and JEOL 3000F transmission electron microscopes. A GATAN liquid
helium holder was used for low temperature TEM measurements. Cr$_{1-x}$V$_{x}$
samples were electropolished for TEM experiments using a methanol:perchloric acid solution (90:10 by volume)at T=260 K.

\subsection{Results and Discussion}

\begin{figure}
\includegraphics[scale=0.45]{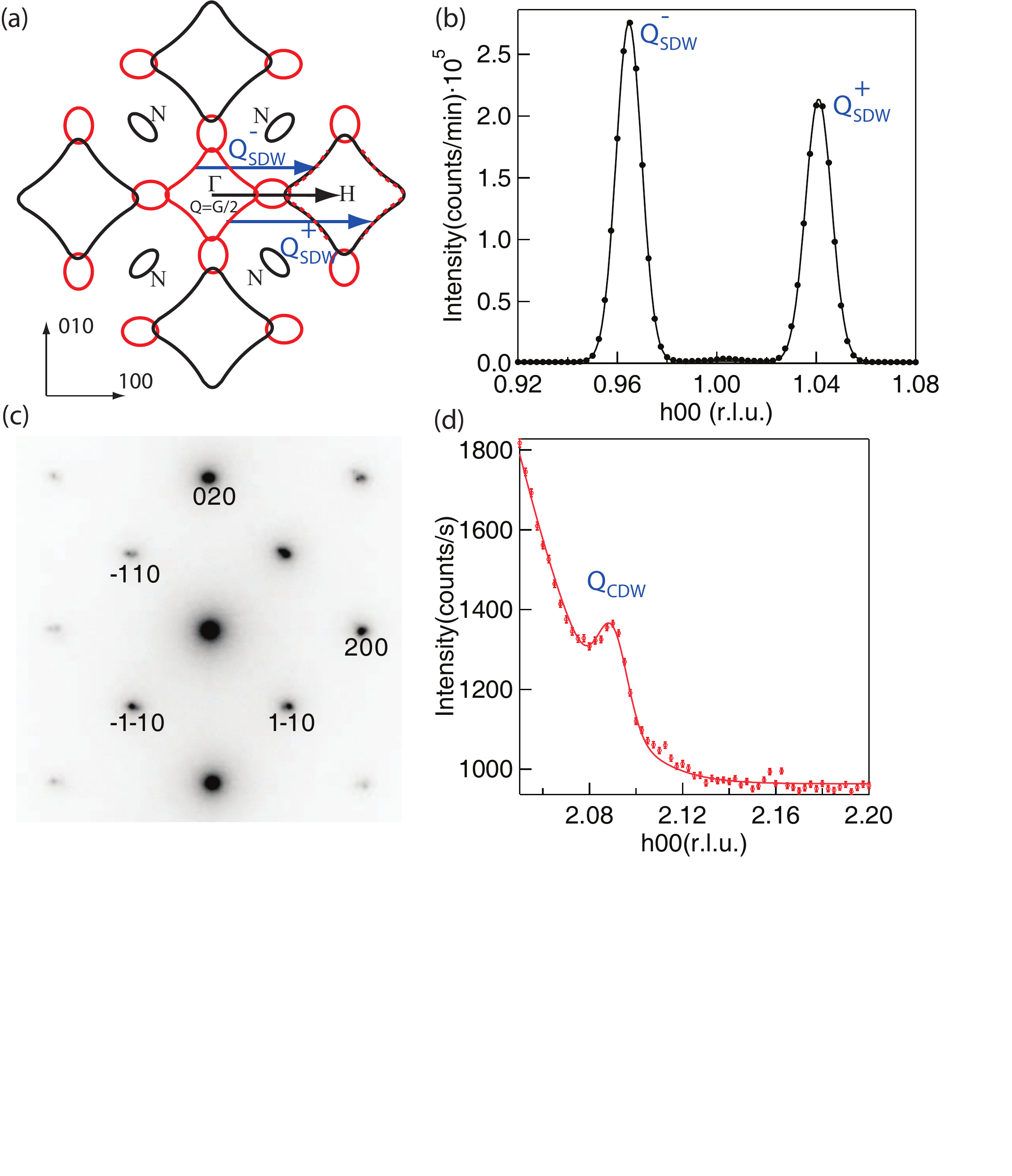}
\caption{\label{fig:epsart} (Color online) Magnetic and Charge Modulation in Cr:  a, Nesting of the $\Gamma$ (electron) and H (hole) FS sheets of Cr by vectors Q$^{+}_{SDW}$ and Q$^{-}_{SDW}$. Electron lenses located along the vertices of the electron octahedron and hole pockets at N do not contribute to static SDW order. b, SDW satellites at Q$^{+}_{SDW}$ and Q$^{-}_{SDW}$ in a 300 K neutron diffraction scan in pure Cr along the [h00] direction near the $\emph{bcc}$ extinct 1,0,0 Bragg peak. The intensities of the peaks at Q$_{SDW}$=2$\pi$/$\emph{a}$(1$\pm$$\delta_{SDW}$,0,0) obey the magnetic form factor, as expected for the Cr SDW~\cite{koehler1966}. c, (h,k,0) plane from TEM experiment on Cr at 300 K, where the apparent elongation of the $\bar{1}$10 Bragg peak arises from incommensurate CDW satellites that are only present below T$_{N}$ at Q$_{CDW}$=2Q$_{SDW}$. d, CDW satellite at Q$^{+}_{SDW}$ in a 300 K X-ray diffraction scan along [H00] near 2,0,0 Bragg peak. The incommensurability of CDW $\delta_{CDW}$=0.090$\pm$0.01 r.l.u., which is close to 2$\delta_{SDW}$=0.082$\pm$0.024 taken from the neutron diffraction measurements in (b). Solid lines in b,d are the best fits to Gaussian line shapes.}
\label{FS}
\end{figure}

The SDW in pure Cr is stabilised by the Coulomb attraction between holes and electrons, reinforced by the strong nesting of large parallel parts of their respective FS~ \cite{overhauser1962,fawcett1988}, FIG\ref{FS}a. In a neutron diffraction experiment, the appearance of satellites, FIG\ref{FS}b near the $\emph{bcc}$ h+k+l=odd peaks marks SDW onset at the Ne{\'{e}}l temperature, T$_{N}$.  Substituting V for Cr shrinks the electron FS relative to the hole FS, making a smaller fraction of the FS nestable~\cite{fawcett1994,koehler1966,komura1967}. The result is a SDW with a smaller moment and lower T$_{N}$. As the mismatch between the hole and electron FS becomes larger, the SDW wave vector Q$_{SDW}$ in the doped Cr$_{1-x}$V$_{x}$ becomes increasingly incommensurate with the lattice.

SDW formation is accompanied by charge modulation, since electron-phonon coupling can drive a lattice distortion that reduces the energy cost of the purely electronic part of the SDW.  This CDW was previously detected in both X-ray and neutron diffraction experiments\cite{tsunoda1974,tsunoda1975,pynn1976,hill1995,jaramillo2009} on pure Cr as additional superlattice peaks with Q$_{CDW}$=2Q$_{SDW}$ that appear with the SDW superlattice peaks at T$_{N}$. Our X-ray diffraction experiments confirmed CDW in Cr with Q$_{CDW}$=2$\pi$/$\emph{a}$(2.09,0,0), FIG\ref{FS}d. We have used transmission electron microscopy (TEM) experiments to study CDW formation in Cr, FIG\ref{FS}c~\cite{prekul1979}, confirming that Q$_{CDW}\simeq$2Q$_{SDW}$ (FIG\ref{FS}c) and that the CDW is present at 300 K but absent at 350 K, as expected given that T$_{N}$=311 K in bulk Cr. Taken together, the neutron, X-ray and TEM experiments confirm that corresponding CDW and SDW modulations are present in pure Cr, arising from the same nesting-driven FS instability.

The aim of the experiments presented here is to determine whether this scenario is still appropriate when the SDW instability is driven to very low temperatures by replacing Cr by nonisoelectronic V in Cr$_{1-x}$V$_{x}$. High pressures suppress the SDW by increasing the breadth of the nested bands\cite{jaramillo2008}, while V doping reduces the volume of the $\Gamma$-centered electron FS, while increasing the volume of the hole FS that is centered at H, FIG\ref{FS}a. Electrical transport and specific heat measurements\cite{takeuchi1980,yeh2002} showed that T$_{N}\rightarrow$0 for a critical V doping x$_{C}$$\simeq$0.04. Only recently\cite{jaramillo2008} have neutron diffraction experiments begun to probe the vicinity of the doping-induced QCP, and our measurements show that both the SDW wavevector and ordered moment continue to decrease smoothly with T$_{N}$ as x$\rightarrow$x$_{C}$.

\begin{figure}
\includegraphics[scale=0.45]{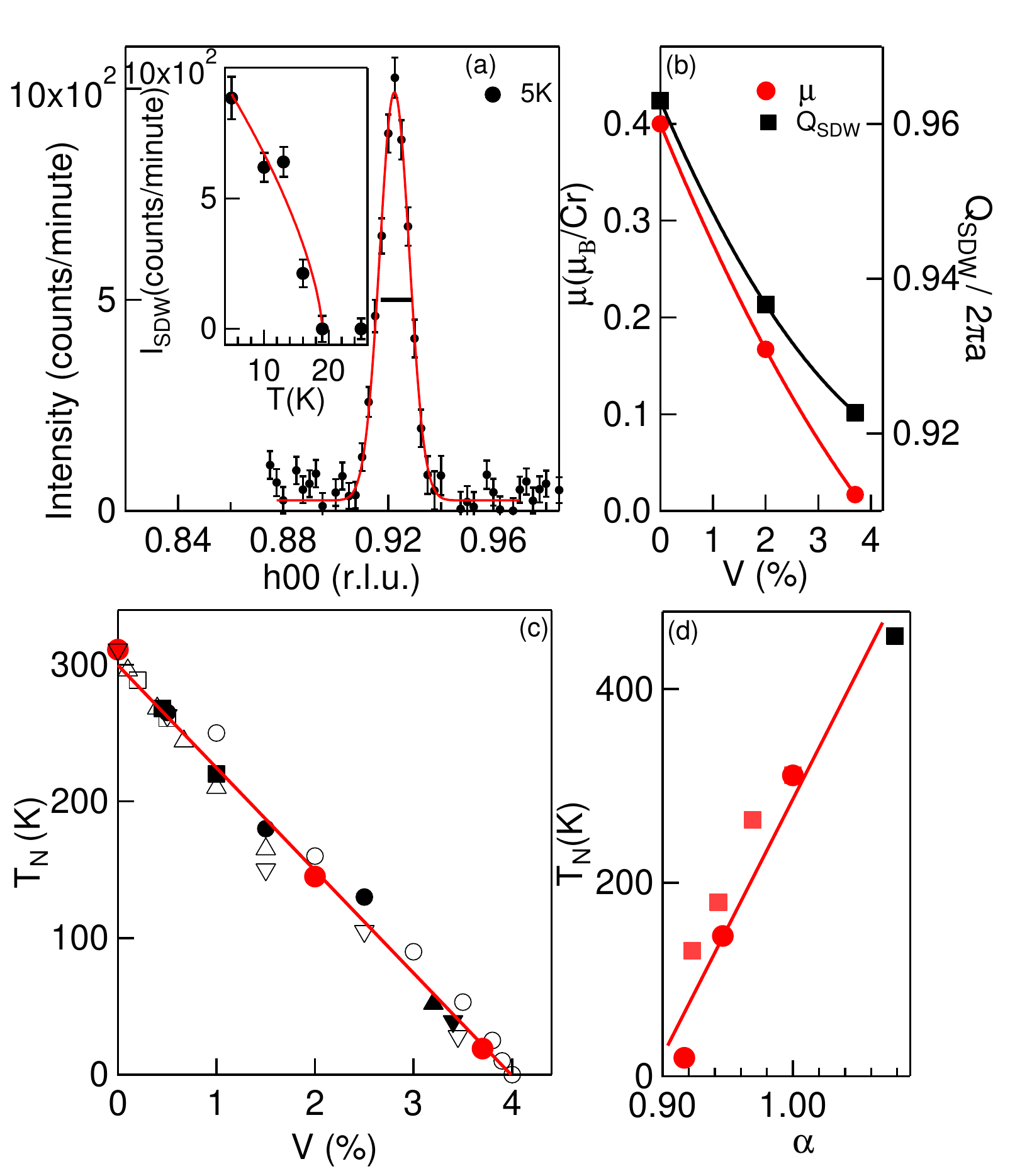}
\caption{\label{fig:epsart} (Color online) Suppression of the SDW in Cr$_{1-x}$V$_{x}$. a, [h00] neutron elastic scan in Cr$_{0.963}$V$_{0.037}$ near the $\emph{bcc}$ extinct (1,0,0) Bragg peak at 5 K. Solid line is best fit to a Gaussian line shape. Inset shows the temperature dependence of the intensity I$_{SDW}$ of the SDW satellite at Q=(0.922,0,0) r.l.u., which vanishes at T$_{N}$=19 K. Solid line is a power law fit.  b, Ordered SDW moments $\mu$ and wave vectors Q$_{SDW}$ for x=0, 0.02, and 0.037. c, Ne\'{e}l temperatures T$_{N}$ for Cr$_{1-x}$V$_{x}$ determined from our measurements (red filled circles) and from ~\cite{takeuchi1980}($\bigcirc$), ~\cite{koehler1966}($\bullet$), ~\cite{komura1967} ($\blacksquare$), ~\cite{noakes1997} ($\square$),~\cite{oliveira1996} ($\vartriangle$), ~\cite{yeh2002}($\blacktriangle$), ~\cite{fawcett1994}($\triangledown$), ~\cite{barnes1965}($\blacktriangledown$). Linear fit has T$_{N}\rightarrow$0 for x$_{C}$=0.040$\pm$0.001. d, Ne\'{e}l temperature as a function of $\alpha$, the ratio of the total FS area nested in Cr$_{1-x}$V$_{x}$, and Cr$_{1-x}$Mn$_{x}$ alloys to the total FS area gapped in pure Cr. Values of T$_{N}$ taken from neutron diffraction measurements (this work: red filled circles), and from more lightly V-doped samples ~\cite{koehler1966} (red squares) and from a Mn-doped sample~\cite{sternlieb1994}($\blacksquare$). Line is a guide for the eye. Error bars in (b-d) are the size of the symbols.}
\label{SDW}
\end{figure}

We have performed neutron diffraction measurements on Cr$_{1-x}$V$_{x}$ with x=0, 0.02, and 0.037, FIG\ref{SDW}. The superlattice peak for the sample with x=0.037 occurs with Q$_{SDW}$=0.922$\pm$0.001 r.l.u., and is only present below the same T$_{N}$=19 K (inset, FIG\ref{SDW}a) found in electrical resistivity measurements~\cite{sokolov2008}, while its breadth indicates that SDW order persists over a length scale that exceeds 34 ${\AA}$. FIG\ref{SDW}b shows that the ordered moment decreases and the SDW becomes increasingly incommensurate as V-doping drives T$_{N}$$\rightarrow$0, consistent with the shrinking of the $\Gamma$ electron FS and the H hole FS with doping represented in FIG\ref{FS}a. Powder and single crystal neutron diffraction and other techniques such as electrical resistivity, thermal expansion, and magnetic susceptibility all agree that T$_{N}$ vanishes approximately linearly at x$_{C}$=0.040$\pm$0.001 (FIG\ref{SDW}c). The amount of nested FS can be modified by electron (Mn) or hole (V) doping, and our results extend earlier measurements~\cite{koehler1966,sternlieb1994} into the QC region (FIG\ref{SDW}d), confirming that a QCP occurs when the electron FS in Cr drops below a critical size.

\begin{figure}
\includegraphics[scale=0.45]{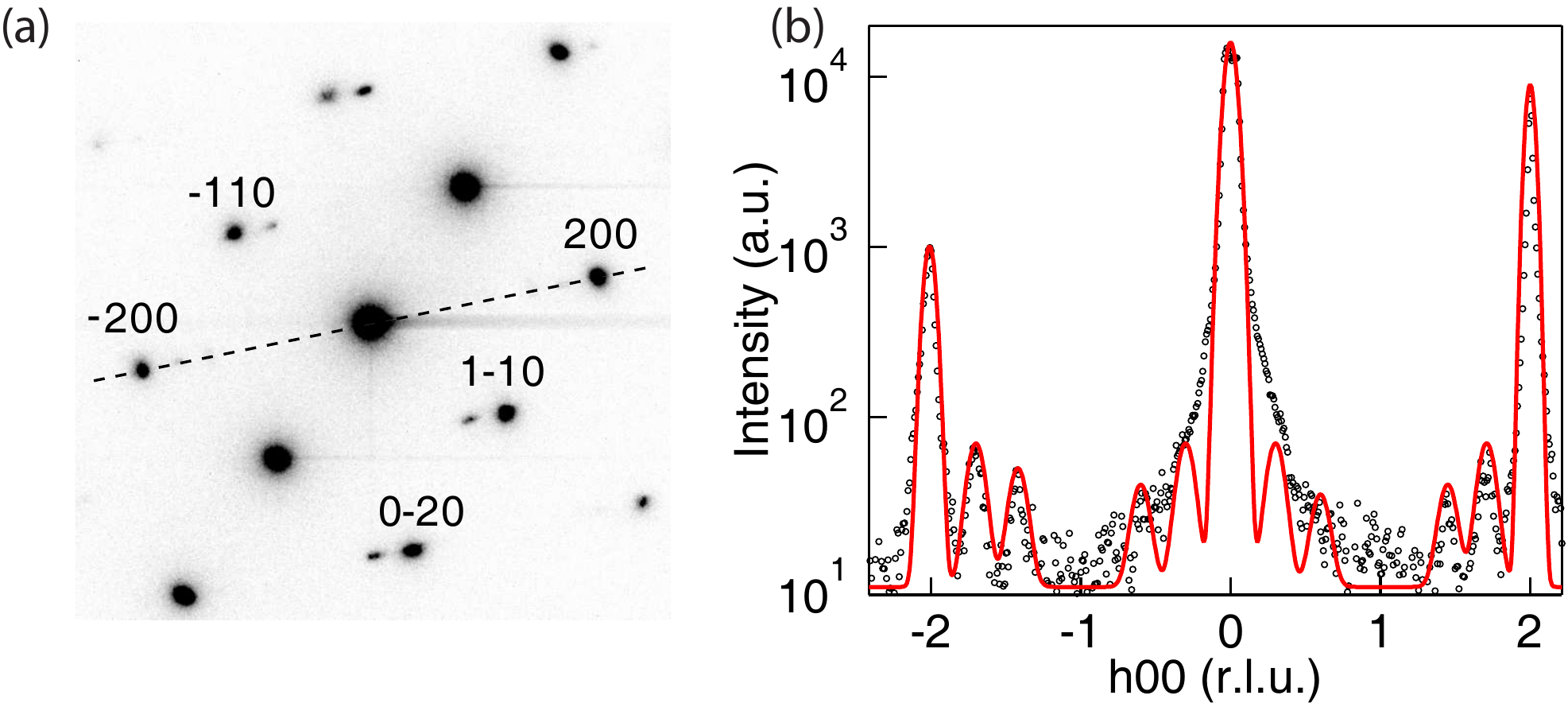}
\caption{\label{fig:epsart} (Color online) Charge Modulation in Cr$_{0.963}$V$_{0.037}$: a, (h,k,0) reciprocal plane at 300 K, where the characteristic splitting of the $\emph{bcc}$ Bragg peaks corresponds to a new charge modulation. Dashed line is the direction of the line cut shown in fig.3b. b, Line cut in [h00] direction in the reciprocal plane, showing the (2,0,0) and ($\overline{2}$,0,0) r. l. u. $\emph{bcc}$ Bragg peaks, each with superlattice peaks with Q$_{SL}$=1.661$\pm$0.001 r. l. u., and their second harmonics with 2Q$_{SL}$. Solid line is a simulation of the line scan, where the lattice Bragg peaks, as well as the first and second harmonics of the charge modulation are modeled as Lorentzian-shaped peaks.}
\label{TEM}
\end{figure}

If FS survives once the SDW is destabilized, and it is interesting to ask whether it persists well into the paramagnetic phase with x$\geq$x$_{C}$ where we have carried out TEM experiments on Cr$_{1-x}$V$_{x}$ (x=0.037). Since we have no $\it{a~priori}\rm$ reason to expect any sort of charge ordering for T$\gg$T$_{N}$, the results are very surprising.  New diffraction peaks are found (FIG\ref{TEM}a) at all temperatures from 88 K - 400 K, corresponding to a superlattice peak with Q$_{SL}$=1.661$\pm$0.001 r.l.u, as well as its second harmonic with 2Q$_{SL}$, indicating that the corresponding charge modulation is not purely sinusoidal, but somewhat flat-topped, FIG\ref{TEM}b.

\begin{figure}
\includegraphics[scale=0.5]{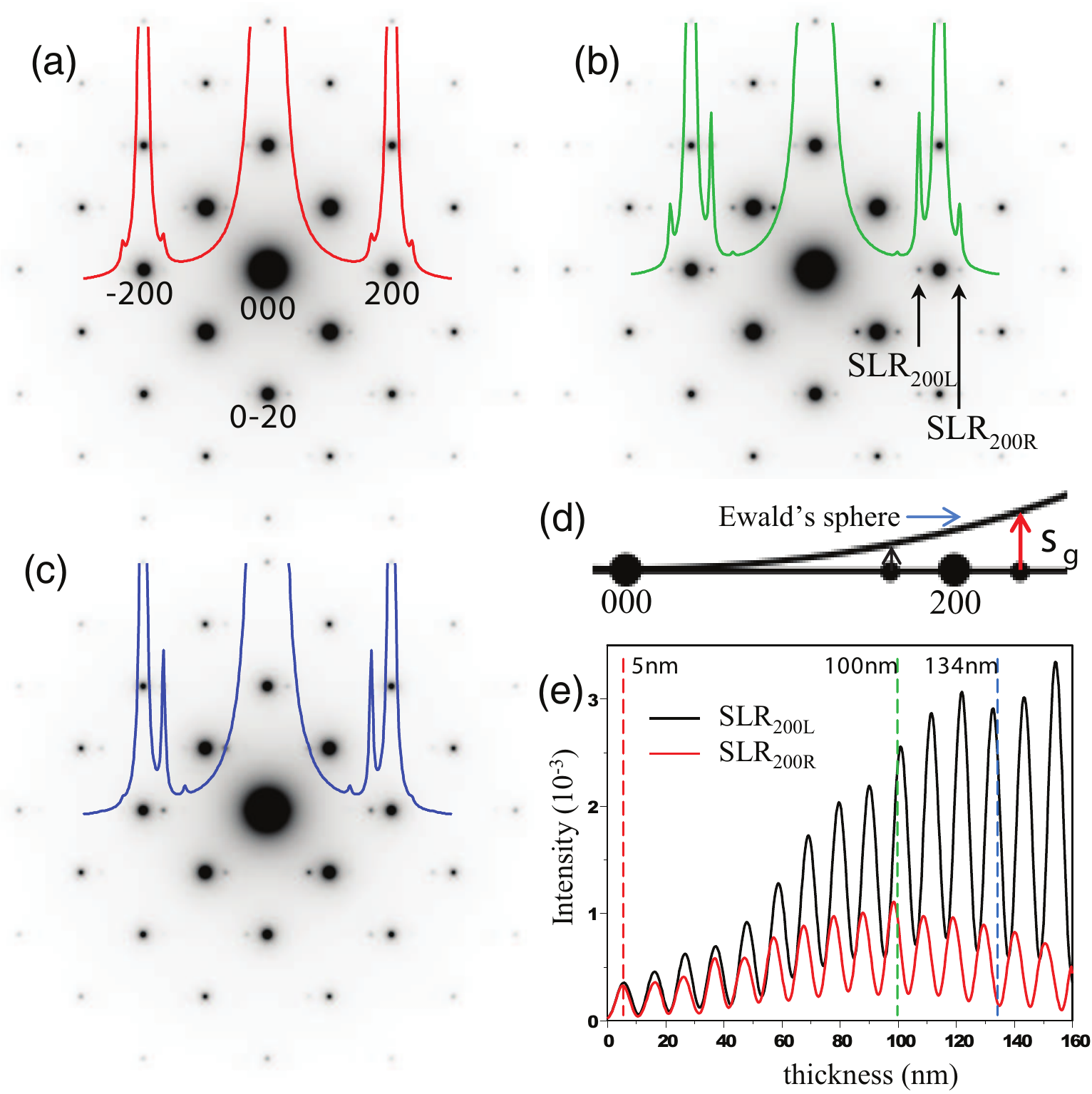}
\caption{\label{fig:epsart} (Color online) (a-c) Calculated [001] electron diffraction patterns with (a) thickness=5 nm, (b) thickness=100 nm and (c) thickness=134 nm. The embedded lines in (a-c) are the intensity line scan from -200 to 200. (d) Schematics of Ewald's sphere in the vicinity of 200 reflection, showing that the excitation error, Sg,  of SLR$_{200L}$ is smaller than that of SLR$_{200R}$. (e) Intensities of SLR$_{200L}$ and SLR$_{200R}$ as a function of thickness.}
\label{CALC_TEM}
\end{figure}

To establish the origin of the missing satellite reflections near $\overline{1}$10, 1$\overline{1}$0 and 0$\overline{2}$0, in FIG\ref{TEM} we have calculated the electron diffraction patterns for samples with increasing thickness using dynamical Bloch wave method for a 3x1x1 superstructure of Cr, FIG\ref{CALC_TEM}. A cosine wave displacement modulation was introduced for superlattice reflections (SLR). The patterns were convoluted with a Lorentzian spread function. At small sample thickness, e.g. in FIG\ref{CALC_TEM}a, the dynamical coupling among the reflections was weak and the electron diffraction was similar to the X-ray diffraction. The intensities of the superlattice reflections SLR at left (SLR$_{200L}$) and right (SLR$_{200R}$) of the 200 reflection were almost the same, though both intensities were very weak. For a thicker sample, e.g. in FIG\ref{CALC_TEM}b, the dynamical effect multiplying the intensity was stronger than in FIG\ref{CALC_TEM}a. Some SLRs are enhanced more strongly than the others due to their different excitation errors or deviations from the Ewald's sphere, resulting in strong asymmetry of the SLRs at left and right of the main reflections FIG\ref{CALC_TEM}d. At large thickness, e.g. in FIG\ref{CALC_TEM}c, the SLR$_{200R}$ was not observed. We conclude that the relative intensities of the SLRs strongly depend on the sample thickness and differ considerably as the thickness increases as shown in FIG\ref{CALC_TEM}e.

\begin{figure}
\includegraphics[scale=0.45]{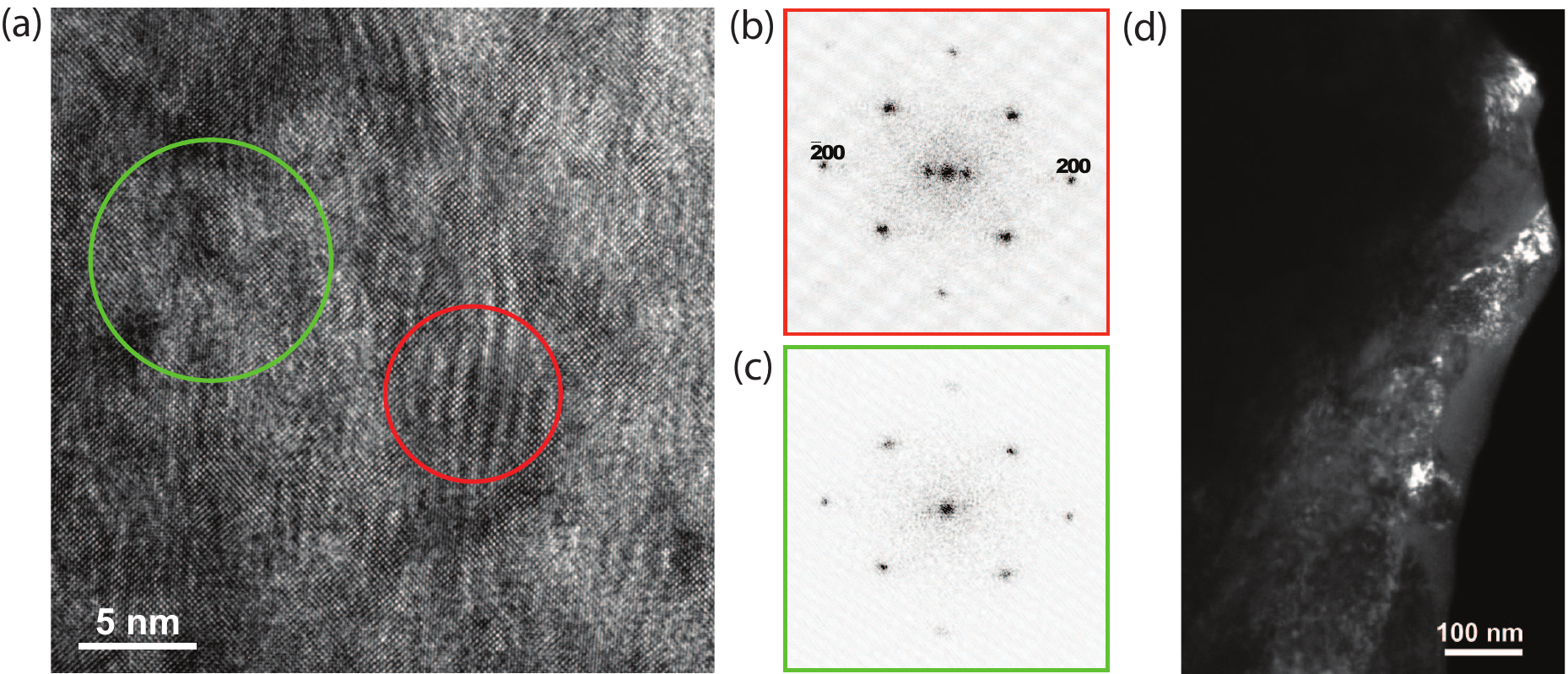}
\caption{\label{fig:epsart} (Color online) Charge Stripes in Quantum Critical Cr$_{0.963}$V$_{0.037}$. a, Direct space image
with atomic resolution showing long vertical fringes that extend over length scales of $\simeq$5 nm. Region inside the red circle has charge stripe modulation, region inside the green circle represents the unmodulated matrix. b, The diffraction patterns obtained from the fast Fourier transform (FFT) of the real space image are different for the stripe regions (red circle) and for the matrix in which they are embedded (green circle). FFT of striped area encircled in red shows 2 incommensurate superlattice satellites near (0,0,0). c, FFT of green encircled area shows no satellites, only the $\emph{bcc}$ Bragg peaks. d, 300 K dark field image where the bright regions indicate parts of the sample with enhanced intensity at Q=Q$_{SL}$.}
\label{STRIPES}
\end{figure}

A real space image of the charge modulation shown in FIG\ref{TEM}a is depicted in FIG\ref{STRIPES}a. The image shows a stripelike modulation with a wave length $\lambda_{SL}\simeq$3$\pm$0.3$\emph{a}$. The SDW modulation in this x=0.037 sample has Q$_{SDW}$=(0.922$\pm$0.001,0,0) r. l. u., and its corresponding CDW would have Q$_{CDW}$=2Q$_{SDW}$=(1.844,0,0) r.l.u., yielding $\lambda_{CDW}\simeq$ 6.4$\emph{a}$, which is about twice the wavelength of the new striped modulation. Both are considerably shorter than the CDW wavelength of $\simeq$13 $\emph{a}$  found in pure Cr at 300 K. This new striped modulation is unrelated to the SDW/CDW found for T$\leq$T$_{N}$=19 K, or from the Cr inclusions found in this sample. Similarly the absence of an SDW with Q$_{SDW}$=0.5 Q$_{SL}$ indicates that the striped instability affects only the charge and not the magnetic moment.

FIG\ref{STRIPES}a shows that the striped phase is not spatially homogeneous, but is restricted to regions whose average size is $\approx$5 nm, where Q$_{SL}$ is virtually identical in every region measured.  The FFT of the striped regions (FIG\ref{STRIPES}b) produced clear incommensurate satellites at the same wave vector Q$_{SL}$ in each region tested. No superlattice peaks were observed in the matrix (FIG\ref{STRIPES}c), which we conclude is simply the paramagnetic metal expected for this composition and temperature. The dark field image (FIG\ref{STRIPES}d) shows that about 20-25 $\%$ of the imaged area is in the striped phase having Q$_{SL}$ parallel to the indicated [h00] direction. Assuming that there is some angular distribution of Q$_{SL}$ that prohibits all striped regions from being imaged simultaneously, the actual volume fraction may be considerably larger.

\begin{figure}
\includegraphics[scale=0.5]{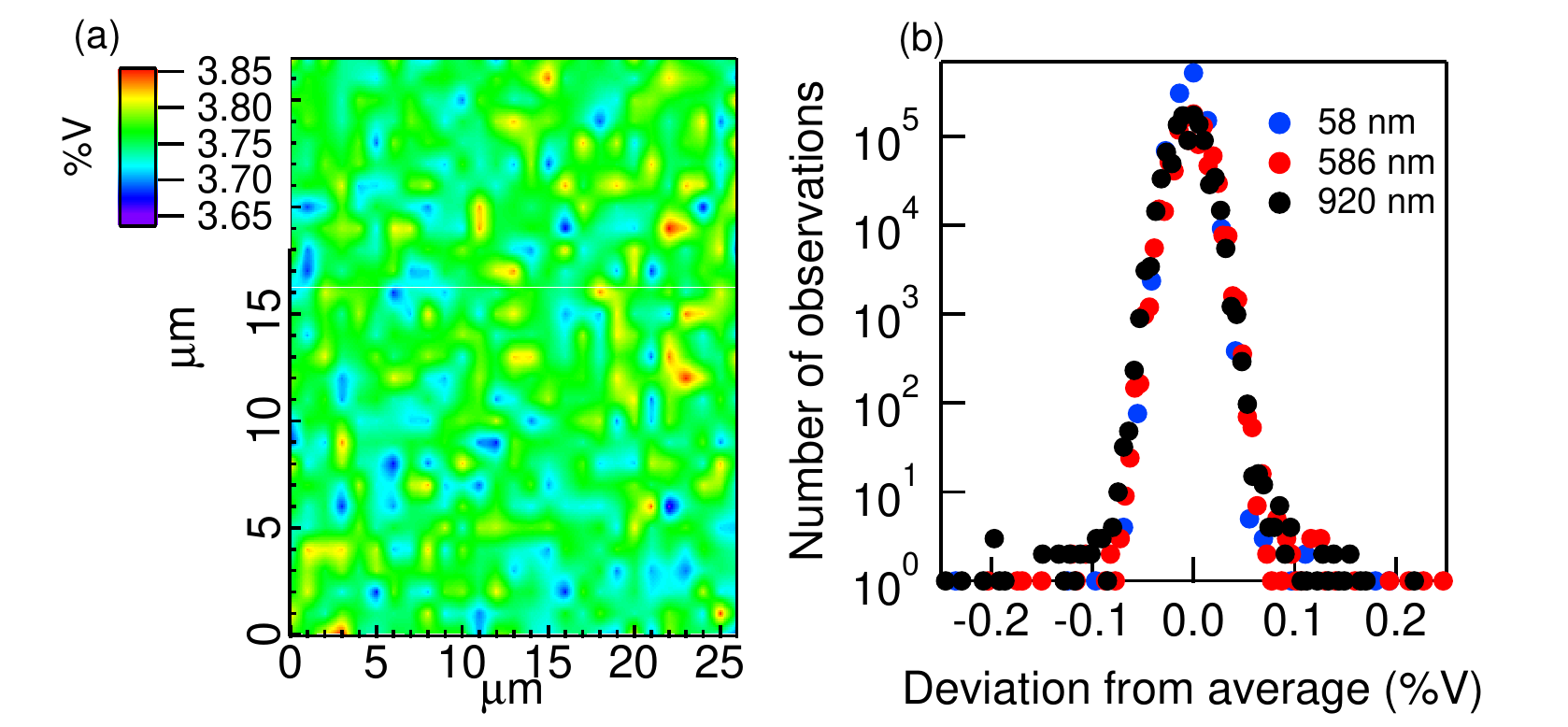}
\caption{(a), Contour plot of the spatial variation of V
concentration obtained from microprobe measurements. The V concentration varies by no more than 4$\%$ over the selected area. (b) Histogram of electron energy loss spectroscopy measurements: number of observations for a given V
concentration versus deviations from the average V concentration
probed on length scales from 58 nm to 920 nm. The average deviation from the mean V concentration is less than 0.1$\%$ for length scales from 58 nm - 920 nm.}
\label{EELS}
\end{figure}

The general morphology revealed in FIG\ref{STRIPES}, where regions with charge order are embedded in a matrix with exemplary crystalline order but no charge superlattice, is reminiscent of the phase separation that accompanies first order transitions. One source may be the spinodal decomposition of a binary alloy that occurs on cooling from the melt, producing regions that are V-deficient and V-rich, relative to the average V concentration x$_{C}$.  In this scenario, the V concentration in the matrix x$_{C}^{-}\leq$x$_{C}$, and so a SDW would be present for T$\leq$T$_{N}$=19 K. Conversely, the absence of a second SDW in the temperature range 19 K$\leq$T$\leq$ 400 K implies that the striped regions would have a V concentration x$_{C}^{+}$ that is larger than x$_{C}$. We have used electron microprobe (FIG\ref{EELS}a) and electron energy loss spectroscopy (FIG\ref{EELS}b) to demonstrate that the V concentration varies by no more than 1000 ppm on any length scale larger than $\approx$58 nm, and the lack of variability in $\lambda_{SL}$ among different striped regions suggests that this is likely also true for length scales that approach the average 5 nm size of the striped regions themselves. Indeed, the binary phase diagram~\cite{villars} implies continuous solubility of V in Cr, arguing against compositional phase separation.  It is more likely~\cite{burgy2001,uemura2006} that the droplets of charge ordered phase in a uniform SDW matrix represent the coexistence region of a first order transition that separates the SDW/CDW found for x$\leq$x$_{C}$ from an instability that affects only the charge that is found for x$\geq$x$_{C}$.

What is the nature of the striped charge modulation that replaces the familiar SDW in Cr for x$\geq$ x$_{C}$? We observe that the magnitude of q$_{SL}$ is very similar to that of a wave vector q$_{SDW,H}$ that connects the parallel surfaces of the H-centered hole FS, suggesting the following scenario. Initially the FS nesting wave vector in pure Cr connects two different sheets of the FS. The number of nested states decreases with V doping, and eventually is not sufficient to make this SDW/CDW energetically favorable when x$\geq$x$_{C}$. The failure of the original SDW/CDW leaves the FS ungapped. The electronic structure calculations for paramagnetic Cr suggest that the FS nesting can occur at multiple vectors either along [011] or [111]~\cite{laurent1981}. Indeed, the inelastic neutron scattering reported multiple Kohn anomalies in the phonon spectrum of pure Cr, which can soften in doped Cr, stabilizing CDW~\cite{shaw1971}. More recently, inelastic X-ray scattering experiments with a superior wavevector resolution found softening of phonons in pure chromium that occur far from the established wavevectors of the Fermi surface~\cite{lamago2010}. The study suggests that the strong electron-phonon coupling alone can drive the phonon softening without invoking Fermi surface nesting. We note, however that q$_{SL}$ is parallel to [h00] in FIG\ref{TEM}a and FIG\ref{STRIPES}b, whereas wavevector of the soft phonon occurs at q=(0.5,0.5,0) suggesting either change of the topology of the electron balls and hole lenses or rotation of the electron jack relative to undoped Cr. Whether charge modulation in Cr$_{0.963}$V$_{0.037}$ originates from hidden Fermi Surface nesting or due to soft phonon remains an open question.

Although the binary phase diagram for Cr-V may argue otherwise, it is even possible that the striped regions form in a new binary compound Cr$_{1000}$V$_{38}$, and that this new chemical periodicity is amplified by nesting to result in the formation of the CDW gap, analogous to the FS-driven superstructures seen in other binary compounds such as Cu-Au or Au-Mn, in which a substantial element substitution stabilizes a new charge modulation ~\cite{tachiki1966}.  Why is this instability, which affects only the charge and not the magnetic moment, absent in pure Cr? It has been suggested~\cite{yeh2002} that it may exist dynamically, as a pseudogap akin to those discussed in more complex systems such as dichalcogenides\cite{borisenko2008} and cuprates\cite{wise2008}, only becoming a static and incommensurate CDW once the SDW is completely suppressed. The most powerful evidence for the electronic transformation of Cr$_{1-x}$V$_{x}$ at the QCP comes from Hall effect measurements\cite{yeh2002,lee2004}, which find that the number of carriers approximately doubles with the demise of the SDW at x$_{C}$, giving a T$\rightarrow$0 Hall number R$_{H}^{-1}$=0.35 electrons/Cr for x$<$x$_{C}$ that is in reasonable agreement with the value R$_{H}$$^{-1}$=0.54 electrons/Cr deduced from  electronic structure calculations\cite{norman2003}, assuming the complete absence of gapping for both $\Gamma$ and H FS.

\subsection{Conclusion}
Our results underscore the powerful role that nesting plays in the different phases of Cr$_{1-x}$V$_{x}$, initially driving the SDW/CDW instabilities familiar from pure Cr, but ultimately replacing them at a critical concentration x$_{C}$ with a new charge instability. While this system shares many of the common features of correlated electron systems with competing orders, such as FS volume changes and the adoption of novel ordered states near the QCP, the perennial value of Cr as a benchmark system is that a detailed and experimentally verified understanding of the electronic structure is available, and our studies of the magnetic and charge modulation advance this understanding into a new regime.

\subsection{Acknowledgements}
DAS and MCA would like to thank A. M. Tsvelik, J. Tranquada, T. M. Rice and S. M. Shapiro
for useful discussions. Work at BNL was carried out under the auspices of the U.S. Department of Energy, Office of Basic Energy Sciences under Contracts No. DE-AC02-98CH1886 (MCA and DAS) and DE-AC02-98CH10886 (YZ and CN). Work at ORNL (ML and SEN) was carried out under the auspices of the U.S. Department of Energy, Office of Basic Energy Sciences, Scientific User Facilities Division. The electron microprobe at the University of Michigan used in this study was partially funded by grant EAR-99-11352 from the National Science Foundation. We acknowledge the support of the National Institute of Standards and Technology, U.S. Department of Commerce in providing the neutron research facilities used in this work. Identification of commercial equipment in the text is not intended to imply recommendation or endorsement by the National Institute of Standards and Technology.

\end{document}